\documentclass[11pt]{article}
\usepackage{amssymb, latexsym, amsfonts, amsmath, amsthm}
\begin{document}

\date{}

\begin{center}
{\Large {\bf MODIFIED TRIPLECTIC QUANTIZATION}}\\
\vspace{.2cm}
{\Large {\bf IN GENERAL COORDINATES}}
\end{center}

{\Large
\begin{center}
\bigskip {\sc B.~Geyer}$^{\ a,}$\footnote{E-mail:
geyer@itp.uni-leipzig.de} and {\sc P.M.~Lavrov}$^{\ a,b,}$\footnote{E-mail:
lavrov@tspu.edu.ru;
lavrov@itp.uni-leipzig.de}\\

\vspace{.5cm} {\normalsize\it $^{a)}$ Institute of Theoretical
Physics and Center of Theoretical Studies,\\ Leipzig University,
%Augustusplatz 10/11,
D-04109 Leipzig, Germany}\\  {\normalsize\it $^{b)}$ Tomsk State
Pedagogical University, 634041 Tomsk, Russia}
\end{center}
}
\vspace{.5cm}

\begin{quotation}
\setlength{\baselineskip}{10pt} {\small \noindent }

{\small We present an extension of the previous results \cite{gln} on the
quantization of general gauge theories within the BRST--antiBRST invariant
Lagrangian scheme in general coordinates. Namely, we generalize \cite{gln}
to the case when the base manifold of fields and antifields is a
supermanifold described in terms of both bosonic and fermionic coordinates.
}
\end{quotation}

\section{Introduction}

%\bigskip\begin{center}{\bf 1. INTRODUCTION}\end{center}

Modern covariant quantization methods for general gauge theories are based
on the principle of BRST \cite{bv,brst}, or, more generally, BRST--antiBRST
\cite{BLT,3pl,anti,mod3pl} invariance. The consideration of these methods in
general coordinates (using appropriate supermanifolds) appears to be very
important in order to reveal the geometrical meaning of the basic objects
underlying these quantization schemes.

The study of the Batalin--Vilkovisky (BV) method \cite{bv} in general
coordinates was initiated by the work of Witten \cite{w}, where the
geometrical meaning of the antifields, the antibracket, and the odd
second-order operator $\Delta $ was discussed. In \cite{geom}, it was shown
that the geometry of the BV formalism is that of an \textit{odd} symplectic
superspace, endowed with a density function $\rho $.

The quantization schemes based on the BRST--antiBRST symmetry involve
additional basic objects. Namely, in the $Sp(2)$-covariant and triplectic
quantization schemes one introduces $Sp(2)$-doublets of extended
antibrackets, as well as doublets of second- and first-order operators $%
\Delta ^{a}$ and $V^{a}$, respectively. In the modified triplectic
quantization, an additional $Sp(2)$-doublet of first-order operators $U^{a}$
is required. This indicates that the geometrical formulation of these
quantization methods in general coordinates, in contrast to the BV
quantization, requires more complicated tools. Indeed, in this paper we show
that the geometry of the $Sp(2)$-covariant and triplectic schemes is the
geometry of an \textit{even} symplectic superspace equipped with a density
function and a flat symmetric connection (covariant derivative), while the
geometry of the modified triplectic quantization also includes a symmetric
structure (analogous to a metric tensor).

The paper is organized as follows. In Sect. 2, we briefly review
the definitions of tensor fields, the covariant derivative, and
the curvature tensor on supermanifolds \cite{DeWitt}. In Sect. 3,
we give the definition of a \emph{triplectic supermanifold,}
together with tensor fields and covariant derivatives acting on
it. In Sect. 4, an explicit realization of the triplectic algebra
of odd differential operators is suggested. In Sect. 5, we find a
realization of the modified triplectic algebra and propose a
suitable quantization procedure. In Sect. 6, we give a short
summary and a few concluding remarks. In Appendix A, we study the
connection between even (odd) non-degenerate Poisson structures
and even (odd) symplectic structures on supermanifolds, and show
their one-to-one correspondence. In Appendix B, the algebra of the
generating operators $\Delta ^{a}$ and $V^{a}$ is presented.

We use the condensed DeWitt notation and apply the tensor analysis of Ref.~
\cite{DeWitt}. Derivatives with respect to the variables $x^{i}$ are
understood as acting from the left, with the notation $\partial _{i}A={%
\partial A}/{\partial x^{i}}$. Right-hand derivatives with respect
to $x^{i}$ are labelled by the subscript $"r"$, and the notation $A_{,i}={%
\partial _{r}A}/{\partial x^{i}}$ is used. Raising the $Sp(2)$-group indices
is performed by the antisymmetric second rank tensor $\varepsilon ^{ab}$ ($%
a,b=1,2$): $\theta ^{a}=\varepsilon ^{ab}\theta _{b}$, $\varepsilon
^{ac}\varepsilon _{cb}=\delta _{b}^{a}$. The Grassmann parity of any
quantity $A$ is denoted by $\epsilon (A)$.

\section{Tensor fields, covariant derivative and curvature tensor on\newline
supermanifolds}

In this Section, we briefly review the tensor analysis on
supermanifolds as far as it is required for the following
considerations. For a comprehensive treatment, we recommend
Ref.~\cite{DeWitt}.

Let the variables $x^{i},\epsilon (x^{i})=\epsilon _{i}$ be local
coordinates of a supermanifold $M$, ${\rm dim}\,M=N$, in the
vicinity of a point $P$. Let the sets $\{e_{i}\}$ and $\{e^{i}\}$
be coordinate bases in the tangent space $T_{P}M$ and the
cotangent space $T_{P}^{\ast }M$, respectively. Under a change of
coordinates, $x^i\; \rightarrow \;{\bar{x}}^{i}={\bar{x}}^{i}(x)$,
the basis vectors in $T_{P}M$ and $T_{P}^{\ast }M$ transform
according to
\begin{eqnarray}
\label{vec}
 {\bar e}_i=e_j \frac{\partial_r x^j}{\partial {\bar x}^i}\,,
 \qquad
 {\bar e}^i=e^j \frac{\partial {\bar x}^i}{\partial x^j}\,.
\end{eqnarray}

Tensor fields of type $(n,m)$ with rank $n+m$ on a supermanifold in some
local coordinate system $(x)=(x^{1},...,x^{N})$ are given by a set of
functions $T_{\;\;\;\;\;\;\;\;\;\;j_{1}...j_{m}}^{i_{1}\ldots i_{n}}(x)$, $%
\;\epsilon (T_{\;\;\;\;\;\;\;\;\;\;j_{1}...j_{m}}^{i_{1}\ldots
i_{n}})=\epsilon (T)+\epsilon _{i_{1}}+\cdot \cdot \cdot +\epsilon
_{i_{n}}+\epsilon _{j_{1}}+\cdot \cdot \cdot +\epsilon _{j_{m}}$, which
transform under a change of coordinates, $x^{i}\rightarrow {\bar{x}}^{i}$,
according to
\begin{eqnarray}
{\bar{T}}_{\;\;\;\;\;\;\;\;\;\;\;j_{1}...j_{m}}^{i_{1}...i_{n}}
&=&T_{\;\;\;\;\;\;\;\;\;\;k_{1}...k_{m}}^{l_{1}...l_{n}}\frac{\partial
_{r}x^{k_{m}}}{\partial {\bar{x}}^{j_{m}}}\cdot \cdot \cdot \frac{\partial
_{r}x^{k_{1}}}{\partial {\bar{x}}^{j_{1}}}\frac{\partial {\bar{x}}^{i_{n}}}{%
\partial x^{l_{n}}}\cdot \cdot \cdot \frac{\partial {\bar{x}}^{i_{1}}}{%
\partial x^{l_{1}}}  \label{tenzor} \\
&&\times (-1)^{\left(\sum\limits_{s=1}^{m-1}\sum\limits_{p=s+1}^{m}
\epsilon_{j_{p}}(\epsilon _{j_{s}}+\epsilon_{k_{s}})+
\sum\limits_{s=1}^{n}\sum\limits_{p=1}^{m}\epsilon_{j_{p}}
(\epsilon _{i_{s}}+\epsilon_{l_{s}})+
\sum\limits_{s=1}^{n-1}\sum\limits_{p=s+1}^{n}\epsilon
_{i_{p}}(\epsilon _{i_{s}}+\epsilon _{l_{s}})\right) }.
\nonumber
\end{eqnarray}

For second-rank tensor fields of type $(2,0),\,(0,2)$ and $(1,1)$, one gets
\begin{eqnarray}
\label{formup}
{\bar T}^{ij}&=&
T^{mn}\frac{\partial {\bar x}^j}{\partial x^n}
\frac{\partial {\bar x}^i}{\partial x^m}
(-1)^{\epsilon_j(\epsilon_i+\epsilon_m)},\\
\label{form}
{\bar T}_{ij}&=&
T_{mn}\frac{\partial_r x^n}{\partial {\bar x}^j}
\frac{\partial_r  x^m}{\partial {\bar x}^i}
(-1)^{\epsilon_j(\epsilon_i+\epsilon_m)},\\
\label{form1} {\bar T}^i_{\;\;j}&=&
 T^m_{\;\;\,n}\frac{\partial_r
x^n}{\partial {\bar x}^j} \frac{\partial {\bar x}^i}{\partial x^m}
(-1)^{\epsilon_j(\epsilon_i+\epsilon_m)},\\
\label{form2}
 {\bar T}_i^{\;\;j}&=& T_m^{\;\;n}
 \frac{\partial {\bar x}^j}{\partial x^n}
 \frac{\partial_r x^m}{\partial {\bar x}^i}\,
  (-1)^{\epsilon_j (\epsilon_i+\epsilon_m)}\,.
\end{eqnarray}
Using DeWitt's index shifting rules \cite{DeWitt},
\begin{eqnarray}
\label{shift} T^{ij}&=&(-1)^{\epsilon(T)\epsilon_i}\, {^iT^j},
\quad
T_{ij}=(-1)^{(\epsilon(T)+1)\epsilon_i}\, {_iT_j},\\
T^{i}_{\;\;j}&=&(-1)^{\epsilon(T)\epsilon_i}\, {^iT_j}, \quad
T_i^{\;\;j}=(-1)^{(\epsilon(T)+1)\epsilon_i}\, {_iT^j},
%\\ T^{\;\;i}_j&=&(-1)^{\epsilon_i\epsilon_j}\, ,
\end{eqnarray}
one can rewrite (\ref{formup})--(\ref{form1}) as follows:
\begin{eqnarray}
\label{formsh} {^i{\bar T}^j}&=& \frac{\partial_r {\bar
x}^i}{\partial x^m}\,{^mT^n}\,
\frac{\partial {\bar x}^j}{\partial x^n},\\
\label{formsh1} {_i{\bar T}_j}&=& \frac{\partial x^m}{\partial
{\bar x}^i}\,{_mT_n}\,
\frac{\partial_r  x^n}{\partial {\bar x}^j},\\
\label{formsh2} {^i{\bar T}_j}&=& \frac{\partial_r {\bar
x}^i}{\partial x^m}\,{^m T_n}\,
\frac{\partial_r x^n}{\partial {\bar x}^j},\\
\label{formsh3} {_i{\bar T}^j}&=& \frac{\partial x^m}{\partial
{\bar x}^i}\,{_m T^n}\, \frac{\partial {\bar x}^j}{\partial x^n}.
\end{eqnarray}

The unit matrix $\delta _{j}^{i}$ is related to two tensor fields of type $%
(1,1)$, ${^{i}\delta _{j}}$ and ${_{j}\delta ^{i}}$, according to
\begin{eqnarray}
\label{unit} \delta^i_j={^i\delta_j}=\delta^i_{\;\;j}=
(-1)^{\epsilon_i(\epsilon_j+1)}{_j\delta^i} = {_j\delta^i}.
\end{eqnarray}
In the last equality, we have used the fact that $(-1)^{\epsilon
_{i}(\epsilon _{j}+1)}=1$ iff $i=j$, so that
\begin{eqnarray}
\label{unitJ}
 \frac{\partial_r {\bar x}^i}{\partial x^k}
\frac{\partial_r x^k}{\partial {\bar x}^j}
 ={^i\delta_j}= \delta^i_j\,,
 \qquad % \label{unitj}
 \frac{\partial x^k}{\partial {\bar x}^j}
 \frac{\partial {\bar x}^i}{\partial x^k}
={_j\delta^i}=\delta^i_j\,.
\end{eqnarray}

Obviously, using a tensor field of type $(2,0)$,
${\stackrel{1}{^{i}T^{k}}}$, and a tensor field of type $(0,2)$,
${\stackrel{2}{_{i}T_{k}}}$, one can construct two tensor fields
of type $(1,1)$,
\begin{eqnarray} \label{contr}
{\overset{1}{^iT^k}}{\overset{2}{_kT_j}},\qquad
{\overset{2}{_iT_k}}{\overset{1}{^kT^j}},
\end{eqnarray}
transforming according to (\ref{formsh2}) and (\ref{formsh3}), respectively.
Using a vector field $X^{i}$ and a covector field $P_{i}$, one can construct
a scalar field, according to
\begin{eqnarray}
\label{sc} X^i\,_iP = (-1)^{\epsilon_i(\epsilon(P)+1)}X^i\,P_i.
\end{eqnarray}

By analogy with tensor analysis on manifolds, on supermanifolds one
introduces the covariant derivative\footnote{%
In order to avoid cumbersome notation for signs, we use the \emph{convention}
that the covariant derivation always acts from the right. If necessary, we
denote this by an arrow pointing to the left.}, $\overleftarrow{\nabla }%
\equiv \nabla $, of tensor fields by the requirement that this
operation
should map a tensor field of type $(n,m)$ into a tensor field of type $%
(n,m+1)$, and that, in cases when one can introduce Cartesian
coordinates, it should reduce to the usual partial derivative.

Now, let $x^{i}$ be Cartesian coordinates, and ${\bar{x}}^{i}$ be
arbitrary coordinates. Let us consider a vector field $X^{i}$.
Then, in the coordinate
system $(x)$ we have %(covariant derivative acts on the right)
\begin{eqnarray}
\nonumber X^i\,{\nabla}_j=X^i_{\;\;,j}\,.
\end{eqnarray}
When going over to the system $({\bar{x}})$, by virtue of (\ref{form1}), the
following relation holds:
\begin{eqnarray}
\nonumber {\bar X}^i{\bar \nabla}_j=
    X^m_{\;\;\,,n}\frac{\partial_r x^n}{\partial {\bar x}^j}
    \frac{\partial {\bar x}^i}{\partial x^m}
    (-1)^{\epsilon_j(\epsilon_i+\epsilon_m)}.
\end{eqnarray}
This implies
\begin{eqnarray}
\nonumber {\bar X}^i{\bar \nabla}_j= {\bar X}^i_{\;\;,j} + {\bar
X}^k\,\Gamma^i_{\;\;kj} (-1)^{\epsilon_k(\epsilon_i+1)},
\end{eqnarray}
where $\Gamma _{\;\;jk}^{i}$ are (generalized) Christoffel symbols in the
superspace,
\begin{eqnarray}
\label{Cris}
\Gamma^i_{\;\;kj}=
\frac{\partial_r {\bar x}^i}{\partial x^m}
\frac{\partial^2_r x^m}{\partial {\bar x}^k\partial {\bar x^j}},
\end{eqnarray}
which possess the property of generalized symmetry:
\begin{eqnarray}
\label{Crisp}
\Gamma^i_{\;\;jk}= (-1)^{\epsilon_j\epsilon_k}\Gamma^i_{\;\;kj}\,.
\end{eqnarray}
Similarly, the action of the covariant derivative on a covector
field $P_{i}$ of type $(0,1)$ is given by
\begin{eqnarray}
\nonumber {\bar P}_i{\bar \nabla}_j= {\bar P}_{i,j} +{\bar
P}_k{\tilde\Gamma}^k_{\;\;ij}\,,
\end{eqnarray}
with the notation
\begin{eqnarray}
\label{Crisp'}
{\tilde\Gamma}^i_{\;\;jm}= (-1)^{\epsilon_m(\epsilon_j+\epsilon_k)}
\frac{\partial^2_r {\bar x}^i}{\partial x^k\partial x^n}
\frac{\partial_r x^n}{\partial {\bar x}^m}
\frac{\partial_r x^k}{\partial {\bar x}^j}\,.
\end{eqnarray}
Using the derivation of (\ref{unitJ}), one readily establishes the fact that
\begin{eqnarray}
\nonumber
{\tilde\Gamma}^i_{\;\;jk}=-\Gamma^i_{\;\;jk}\,,
\end{eqnarray}
and therefore
\begin{eqnarray}
\nonumber {\bar P}_i{\bar \nabla}_j= {\bar P}_{i,j} -{\bar
P}_k\Gamma^k_{\;\;ij}\,.
\end{eqnarray}

The action of the covariant derivative on second-rank tensor fields of type $%
(2,0)$, $(0,2)$ and $(1,1)$ can be deduced in the same manner as follows:
\begin{eqnarray}
%\nonumber
{\bar T}^{ij}{\bar \nabla}_k&=&
{\bar T}^{ij}_{\;\;\;,k} +
{\bar T}^{im}\Gamma^j_{\;\;mk}(-1)^{\epsilon_m(\epsilon_j+1)}+
{\bar T}^{mj}\Gamma^i_{\;\;mk}
(-1)^{\epsilon_m(\epsilon_i+\epsilon_j+1)+\epsilon_i\epsilon_j},\\
%\nonumber
{\bar T}_{ij}{\bar \nabla}_k&=&
{\bar T}_{ij,k}
-{\bar T}_{il}\Gamma^l_{\;\;jk}-
{\bar T}_{lj}\Gamma^l_{\;\;ik}
(-1)^{\epsilon_j(\epsilon_i+\epsilon_l)},\\
%\nonumber
{\bar T}^i_{\;\;j}{\bar \nabla}_k&=&
{\bar T}^i_{\;\;j,k} -
{\bar T}^i_{\;\;l}\Gamma^l_{\;\;jk}
+{\bar T}^l_{\;\;j}\Gamma^i_{\;\;lk}
(-1)^{\epsilon_i\epsilon_j+\epsilon_l(\epsilon_i+\epsilon_j+1)}.
%\nonumber
\end{eqnarray}

Similarly, one determines the action of the covariant derivative on a tensor
field of any rank and type in terms of the tensor components, ordinary
derivatives and Christoffel symbols. Of course, on arbitrary supermanifolds $%
M$ the Christoffel symbols (i.e. connection coefficients) are not
necessarily given by second-order partial derivatives with respect to the
coordinates, since such a simple form arises only when local Euclidean
coordinates can be introduced on $M$. If one chooses a coordinate system on
a supermanifold, then the covariant derivative $\overleftarrow{\nabla }=(%
\overleftarrow{\nabla }_{i})$ is defined as a variety of
differentiations with respect to separate coordinates. These
differentiations are local operations acting on a scalar field $S$
by the rule
\begin{eqnarray}
\label{scal} S\nabla_i=S_{,i},
\end{eqnarray}
on a vector field $X^{i}$, by the rule
\begin{eqnarray}
\label{vector} X^i\nabla_j=X^i_{\;\;,j}+ X^k\Gamma^i_{\;\;kj}
(-1)^{\epsilon_k(\epsilon_i+1)},
\end{eqnarray}
on a covector field $P_{i}$, by the rule
\begin{eqnarray}
P_i\nabla_j=P_{i,j}-P_k\Gamma^k_{\;\;ij},
\end{eqnarray}
and so on. If the Christoffel symbols are symmetric ones (\ref{Crisp}), then
one says that on the supermanifold $M$ a symmetric connection is defined.
Here, we consider the case of symmetric connections only.

The curvature tensor $R_{\;\;mjk}^{i}$ is defined in a coordinate
basis \cite {DeWitt} by the action of the commutator of covariant
derivatives $[\nabla _{i},\nabla _{j}]=\nabla _{i}\nabla
_{j}-(-1)^{\epsilon _{i}\epsilon _{j}}\nabla _{j}\nabla _{i}$ on a
vector field $X^{i}$ by the rule
\begin{eqnarray}
X^i[\nabla_j,\nabla_k]=-(-1)^{\epsilon_m(\epsilon_i+1)}
X^mR^i_{\;\;mjk}\,.
\end{eqnarray}
A straightforward calculation yields the following result:
\begin{eqnarray}
\label{R}
R^i_{\;\;mjk}=-\Gamma^i_{\;\;mj,k}+
\Gamma^i_{\;\;mk,j}(-1)^{\epsilon_j\epsilon_k}+
\Gamma^i_{\;\;js}\Gamma^s_{\;\;mk}(-1)^{\epsilon_j\epsilon_m}-
\Gamma^i_{\;\;ks}\Gamma^s_{\;\;mj}
(-1)^{\epsilon_k(\epsilon_m+\epsilon_j)}.
\end{eqnarray}
The curvature tensor (\ref{R}) obeys the following generalized
antisymmetry:
\begin{eqnarray}
\label{Rsym}
R^i_{\;\;mjk}=-(-1)^{\epsilon_j\epsilon_k}R^i_{\;\;mkj}\,.
\end{eqnarray}
One readily establishes the fact that the curvature tensor
(\ref{R}) also obeys the Jacobi identity
\begin{eqnarray}
\label{Rjac}
(-1)^{\epsilon_m\epsilon_k}R^i_{\;\;mjk}
+(-1)^{\epsilon_j\epsilon_m}R^i_{\;\;jkm}
+(-1)^{\epsilon_k\epsilon_j}R^i_{\;\;kmj}=0.
\end{eqnarray}

\section{Triplectic supermanifolds}

The supermanifolds arising within the triplectic \cite{3pl} and modified
triplectic \cite{mod3pl} quantization schemes can be identified with
superspaces parameterized by the variables $x^{i}=(\phi ^{A},\bar{\phi}_{A})$,
$i=1,2,\ldots,N=2n$, where $\phi ^{A}$ are the (field) variables of the
configuration space of a general gauge theory, and where the antifields $%
\bar{\phi}_{A}$ are the sources of the combined BRST--antiBRST symmetry. The
complete superspace of these quantization methods also involves variables $%
\theta _{ia}=(\phi _{Aa}^{\ast },\pi ^{Aa})$ with Grassmann parity $\epsilon
(\theta _{ia})=\epsilon _{i}+1$, opposite to $x^{i}$. Here, the antifields $%
\phi _{Aa}^{\ast }$ are the sources of the BRST ($a=1$) and antiBRST ($a=2$)
transformations, while the fields $\pi ^{Aa}$ are auxiliary ones. In the
original formulation of $Sp(2)$-covariant quantization, these variables are
used to introduce the gauge. The character $a$ indicates the (global)
$Sp(2)$
group index. (We remind that in Ref.~\cite{gln} only the case $\varepsilon
_{i}=0$ was considered.)

In order to formulate the modified triplectic quantization in
general coordinates, let us consider a supermanifold
$\mathcal{M}$, ${\rm dim}\, \mathcal{M}=3N$, which can be locally
described by coordinates $z^{\mu }=(x^{i},\theta _{ia})$, with
$\epsilon (x^{i})=\epsilon _{i}$, $\epsilon
(\theta _{ia})=\epsilon _{i}+1$. Let us introduce the transformation law of $%
\theta _{ia}$ under coordinate transformations ($x\rightarrow {\bar{x}}$) in
the base supermanifold $M$, analogous to the transformation of the basis
vectors in the tangent space $T_{P}M$, namely,
\begin{eqnarray}
\label{theta}
{\bar\theta}_{ia} =\theta_{ja}\frac{\partial_r x^j}{\partial {\bar x}^i}.
\end{eqnarray}
Supermanifolds with such a property will be called
\emph{triplectic supermanifolds} (for a different, more general
introduction of triplectic supermanifolds and a detailed
exposition of their properties, see, e.g.,
Ref.~\cite{3pl,gs1,gs2}). Then, right-hand derivatives with
respect to $\theta _{ia}$ transform like the basis vectors of the
cotangent space $T_{P}^{\ast }M$,
\begin{eqnarray}
\label{dertheta} \frac{\overleftarrow{\partial}
}{\partial{\bar\theta}_{ia}} = \frac{\overleftarrow{\partial}
}{\partial\theta_{ja}} \frac{\partial {\bar x}^i}{\partial x^j}.
\end{eqnarray}

According to Sect. 2, a tensor field of type $(n,m)$ and rank $n+m$ is a
geometric object, which, in any local coordinate system $(x,\theta )$ on $%
\mathcal{M}$, is given by a set of functions, $T_{\;\;\;\;\;\;\;\;\;j_{1}%
\ldots j_{m}}^{i_{1}\ldots i_{n}}(x,\theta )$, transforming under
a change of coordinates of the base manifold $M$, $(x)\rightarrow
({\bar{x}})$, according to (\ref{tenzor}).

Covariant differentiation $(\overleftarrow{\nabla })$ of tensor fields can
be introduced using the same arguments as given above. In particular, the
action of the covariant derivative on the simplest tensor fields (scalar,
vector and second-rank ones) is given by the relations
\begin{eqnarray}
\label{simple} S\nabla_i&=&S_{,i}+
\frac{\partial_r S}{\partial\theta_{ma}}\theta_{ka}\Gamma^k_{\;\;mi},\\
X^i\nabla_j&=&X^i_{\;\;,j}+ X^k\Gamma^i_{\;\;kj}
(-1)^{\epsilon_k(\epsilon_i+1)}+
\frac{\partial_r X^i}{\partial\theta_{ma}}\theta_{ka}\Gamma^k_{\;\;mj},\\
\label{tcov} P_i\nabla_j&=&P_{i,j}-P_k\Gamma^k_{\;\;ij}+
\frac{\partial_r P_i}{\partial\theta_{ma}}\theta_{ka}\Gamma^k_{\;\;mj}, \\
\nonumber
T^{ik}\nabla_j&=&
T^{ik}_{\;\;\;,j} +
T^{im}\Gamma^k_{\;\;mj}(-1)^{\epsilon_m(\epsilon_k+1)}+
T^{mk}\Gamma^i_{\;\;mj}
(-1)^{\epsilon_m(\epsilon_i+\epsilon_k+1)+\epsilon_i\epsilon_k}\\
&&+\frac{\partial_r T^{ik}}{\partial\theta_{ma}}
\theta_{na}\Gamma^n_{\;\;mj},\\
%\nonumber
T_{ik}\nabla_j&=&
T_{ik,j} -T_{il}\Gamma^l_{\;\;jk}-
T_{lj}\Gamma^l_{\;\;ik}
(-1)^{\epsilon_j(\epsilon_i+\epsilon_l)}+
\frac{\partial_r T_{ik}}{\partial\theta_{ma}}\theta_{na}\Gamma^n_{\;\;mj},\\
%\nonumber
T^i_{\;\;k}\nabla_j&=&
T^i_{\;\;k,j}-
T^i_{\;\;l}\Gamma^l_{\;\;jk}
+T^l_{\;\;j}\Gamma^i_{\;\;lk}
(-1)^{\epsilon_i\epsilon_j+\epsilon_l(\epsilon_i+\epsilon_j+1)}+
\frac{\partial_r T^i_{\;\;k}}{\partial\theta_{ma}}
\theta_{na}\Gamma^n_{\;\;mj}.
\end{eqnarray}

Since, according to their introduction, the coordinates $x^{i}$ and $\theta
_{ia}$ are independent of each other, (\ref{tcov}) implies that the vectors $%
\theta _{ia}$ are covariantly constant:
\begin{eqnarray}
\label{thecov}
\theta_{ia}\nabla_j=0.
\end{eqnarray}
Furthermore, from these relations it follows that the action of the
commutator of covariant derivatives on a scalar field is given by
\begin{eqnarray}
S[\nabla_i,\nabla_j]=- \frac{\partial_r S}{\partial
\theta_{ma}}\theta_{na}R^n_{\;\;mij},
\end{eqnarray}
where the curvature tensor $R_{\;\;mij}^{n}$ has been defined in (\ref{R}).

\section{Explicit realization of the triplectic algebra}

The triplectic algebra defined on the triplectic supermanifold $\mathcal{M}$
includes two sets of anticommuting and nilpotent operators of second and
first order, $\overleftarrow{\Delta }^{a}$ and $\overleftarrow{V}^{a}$,
respectively, acting from the right, and obeying the following
algebra:
\begin{eqnarray}
&\Delta^{\{a}\Delta^{b\}} =0,& \quad \epsilon(\Delta^a)=1,\label{dd} \\
&V^{\{a}V^{b\}} =0,&\quad \epsilon(V^a)=1, \label{vv}\\
%&\Delta^{\{a} V^{b\}} + V^{\{a}\Delta^{b\}} = 0&
&V^a\Delta^b +\Delta^bV^a =0, &\label{vd}
\end{eqnarray}
where the curly bracket denotes symmetrization with respect to the enclosed
indices $a$ and $b$. Using the odd second-order differential
operators $\Delta ^{a}$, one can introduce a pair of bilinear operations $%
(\;\,,\;)^{a}$ on the triplectic supermanifold $\mathcal{M}$ by the rule
\begin{eqnarray}
\label{ab}
(F,G)^a = (-1)^{\epsilon(G)}(FG)\Delta^a - (-1)^{\epsilon(G)}(F\Delta^a) G
 - F(G\Delta^a ).
%\label{liebdelta}
\end{eqnarray}
The operations (\ref{ab}) possess the Grassmann parity $\epsilon
((F,G))=\epsilon (F)+\epsilon (G)+1$ and obey the following symmetry
property:
\begin{eqnarray}
\label{symab}
(F,G)^a =-(-1)^{(\epsilon(G)+1)(\epsilon(F)+1)}(G,F)^a.
%\label{liebdelta}
\end{eqnarray}
They are linear operations with respect to both arguments (see (\ref{A3})),
and obey the Leibniz rule (see (\ref{A4})). Due to the properties (\ref{dd})
of the operators $\Delta ^{a}$, these odd bracket operations satisfy the
generalized Jacobi identity
\begin{eqnarray}
\label{jiab}
(F,(G,H)^{\{a})^{b\}}(-1)^{(\epsilon(F)+1)(\epsilon(H)+1)}
+ {\rm cycle} (F,G,H)\equiv 0.
\end{eqnarray}
According to their properties, the operations $(\,\,,\;)^{a}$ form a set of
antibrackets, as have been introduced for the first time in Ref. \cite{BLT}.
Therefore, if we have an explicit realization of the operators $\Delta ^{a}$
with the properties (\ref{dd}), then, according to (\ref{ab}), we can
generate the extended antibrackets explicitly. Explicit realizations of $%
\Delta ^{a}$ are known in two cases: in Darboux coordinates \cite
{BLT,3pl,mod3pl} and in general coordinates when $M$ is a flat Fedosov
manifold, with bosonic variables $x^{i}$ \cite{gln}. However, in Quantum
Gauge Field Theory the base manifold $M$ always requires fermionic
coordinates for its description, and therefore it should be considered as a
supermanifold from the beginning.

Now, we like to extend the considerations of Ref.~\cite{3pl} such
that we not only get some possible explicit realizations of the
triplectic algebra (\ref {dd}) -- (\ref{vd}) on a triplectic
supermanifold $\mathcal{M}$ but, in the next Section, will be able
to extend this to an explicit realization of the modified
triplectic algebra, too.

First, let us equip the base supermanifold $M$ with a Poisson
structure, i.e., with a non-degenerate \emph{even} second-rank
tensor field $\omega ^{ij}(x)$, $\epsilon (\omega ^{ij})=\epsilon
_{i}+\epsilon _{j}$, obeying the property of generalized
antisymmetry,
\begin{eqnarray}
\label{symomega}
\omega^{ij}=-(-1)^{\epsilon_i\epsilon_j}\omega^{ji},
\end{eqnarray}
and satisfying the following identities:
\begin{eqnarray}
\label{jiomega} \omega^{il}\partial_l\omega^{jk}
(-1)^{\epsilon_i\epsilon_k} + {\rm cycle} (i,j,k)\equiv 0.
\end{eqnarray}
This tensor field $\omega ^{ij}$ defines a Poisson bracket, and, due to its
non-degeneracy, also a corresponding even symplectic structure on the
supermanifold $M$ (see Appendix A). At this level, the supermanifold $M$ can
be considered as either an even Poisson supermanifold or an even symplectic
supermanifold.

The Poisson structure $\omega ^{ij}$ allows one to equip the
triplectic supermanifold $\mathcal{M}$ with an
$Sp(2)$--irreducible second-rank tensor $ S_{ab} $,
\begin{eqnarray}
\label{Sab}
S_{ab}=\hbox{\Large$\frac{1}{6}$}\theta_{ia}\omega^{ij}\theta_{jb},
\quad \epsilon(S_{ab})=0,
\end{eqnarray}
which is invariant under changes of the local coordinates on $\mathcal{M}$,
i.e., ${\bar{S}}_{ab}=S_{ab}$, and is symmetric with respect to the $Sp(2)$
indices, $S_{ab}=S_{ba}$.

Now, let us require that the covariant derivative $\nabla _{i}$
should respect the Poisson structure $\omega ^{ij}$:
\begin{eqnarray}
\label{covom} \omega^{ij}\nabla_k=0 \;\; \leftrightarrow \;\;
\omega^{ij}_{\;\;\;,k}+\omega^{im}
\Gamma^j_{\;\;mk}(-1)^{\epsilon_m(\epsilon_j+1)}-
\omega^{jm}\Gamma^i_{\;\;mk}
(-1)^{\epsilon_i\epsilon_j+\epsilon_m(\epsilon_i+1)}=0.
\end{eqnarray}
From (\ref{thecov}) and (\ref{covom}), it follows that the $Sp(2)$ second-rank
tensor $S_{ab}$ is a covariantly constant,
\begin{eqnarray}
\label{conSab}
S_{ab}\nabla_i=0.
\end{eqnarray}

In terms of the tensor field $\omega _{ij}$, the inverse of
$\omega ^{ij}$, the relations (\ref{covom}) read
\begin{eqnarray}
\label{covomiv} \omega_{ij,k}-\omega_{im}\Gamma^m_{\;\;jk}+
\omega_{jm}\Gamma^m_{\;\;ik}(-1)^{\epsilon_i\epsilon_j}=0,
\end{eqnarray}
which provides the covariant constancy of the differential
two-form $\omega $ (see, eq.~\ref{A13}),
\begin{eqnarray}
\label{covform}
\omega\nabla=0.
\end{eqnarray}
Then the supermanifold $M$ can be considered as an even symmetric
symplectic supermanifold, being a supersymmetric extention of the
Fedosov manifold \cite {F,fm}.

Taking into account the relations (\ref{sc}) and (\ref{dertheta}), and using
the covariant operators $\overleftarrow{\nabla _{i}}$ and $\overleftarrow{%
\partial}/\partial \theta _{ia}$, we find that there exists a
unique (up to first-order differential operators) $Sp(2)$-doublet of odd
second-order differential operators acting as scalars on triplectic
supermanifolds $\mathcal{M}$,
\begin{eqnarray}
    \label{Delt}
    \overleftarrow{\Delta^a}=(-1)^{\epsilon_i}
    \frac{\overleftarrow{\partial}}{\partial\theta_{ia}}
    \overleftarrow{\nabla_i}+
    \hbox{\Large$\frac{1}{2}$}(-1)^{\epsilon_i}
    \frac{\overleftarrow{\partial}}{\partial\theta_{ia}}\rho_{,i}\,,
\end{eqnarray}
where $\rho =\rho (x)$,\ $\epsilon (\rho )=0$, is a scalar density on $M$.
The operators (\ref{Delt}) generate a doublet of operations on $\mathcal{M}$,
\begin{eqnarray}
\label{2op} (F,G)^a=(F{\nabla_i})\frac{\partial
G}{\partial\theta_{ia}}-
(-1)^{(\epsilon(F)+1)(\epsilon(G)+1)}(G{\nabla_i}) \frac{\partial
F}{\partial\theta_{ia}}.
\end{eqnarray}
These operations obviously obey all the properties of extended antibrackets,
except for the Jacobi identity, which is closely related to the
anticommutativity and nilpotency (\ref{dd}) of $\Delta ^{a}$.

In order to get also that missing property we must restrict the
base supermanifold somewhat.  Using the operations (\ref{2op}) and the
irreducible second-rank $Sp(2)$-tensor $S_{ab}$ (\ref{Sab}), we
shall define the following $Sp(2)$-doublet of odd first-order
differential operators $\overleftarrow{V_a}$:
\begin{eqnarray}
\label{Va} \overleftarrow{V_a}=(\cdot
\;,S_{ab})^b=\hbox{\Large$\frac{1}{2}$}\overleftarrow{\nabla_i}\,\omega^{ij}\theta_{ja},
\end{eqnarray}
where the relations (\ref{conSab}) have been used for the second equality.
Straightforward calculations, with allowance for the algebra of operators $%
\Delta ^{a}$ (\ref{Delt}) and $V^{a}$ (\ref{Va}) (see Appendix B), show that
there exists a choice of the density function $\rho $, namely,
\begin{eqnarray}
\label{rho}
\rho=-{\rm log}\;{\rm sdet}\;[\omega^{ij}],
\end{eqnarray}
such that the triplectic algebra (\ref{dd}) -- (\ref{vd}) is fulfilled on $\mathcal{M%
}$ when the base supermanifold $M$ is a (flat) Fedosov superspace,
\begin{eqnarray}
\label{R=0} R^i_{\;\;mjk}=0,
\end{eqnarray}
with the curvature tensor $R_{\;\;mjk}^{i}$ given by (\ref{R}). Therefore,
we have explicitly realized the extended antibrackets (\ref{2op}) and the
triplectic algebra (\ref{dd}) -- (\ref{vd}) of the generating operators $%
\Delta ^{a}$, $\overleftarrow{V^{a}}$. Moreover, the operators
$V^{a}$ can always be considered as anti-Hamiltonian vector
fields. Note, that an explicit realization of the antibrackets in
the form (\ref{2op}) for a flat symmetric connection already has
been found in Ref.~\cite{gs2}.

\section{Realization of modified triplectic algebra and quantization}

The modified triplectic quantization \cite{mod3pl}, in comparison
with the $Sp(2)$-covariant method \cite{BLT}, or the triplectic
scheme \cite{3pl}, involves an additional $Sp(2)$-doublet of odd
operators $\overleftarrow{U^{a}}$ $(\epsilon (U^{a})=1)$, with the
following properties:
\begin{eqnarray}
\label{mal}
U^{\{a}U^{b\}}=0,\quad \Delta^{\{a} U^{b\}} + U^{\{a}\Delta^{b\}} = 0,
\quad U^{\{a}V^{b\}} +V^{\{b}U^{a\}}=0.
\end{eqnarray}
An invariant realization of these operators on $\mathcal{M}$
requires to introduce a new geometrical structure on $M$. Namely,
because $M$ is already equipped with the symplectic structure
$\omega _{ij}$, there exists the possibility to equip the base
supermanifold $M$ also with a symmetric second-rank tensor field
$g_{ij}(x)=(-1)^{\epsilon _{i}\epsilon _{j}}g_{ji}(x)$ of type
$(0,2)$.

Notice that on the triplectic supermanifold $\mathcal{M}$ there
exists a vector field $\theta _{a}^{i}$,
\begin{eqnarray}
\label{vom} \theta^i_a=\omega^{ij}\theta_{ja}(-1)^{\epsilon_i},
\end{eqnarray}
which, due to (\ref{thecov}) and (\ref{covom}), is covariantly constant,
\begin{eqnarray}
\label{covthe} \theta^i_a\nabla_j=0.
\end{eqnarray}
This vector field can be used to construct on $\mathcal{M}$ an
$Sp(2)$-scalar function
$S_{0}$, the so-called anti-Hamiltonian, according to
\begin{eqnarray}
\label{S0}
 S_0=\hbox{\Large$\frac{1}{2}$}\varepsilon^{ab}\theta^i_a\;g_{ij}\theta^j_b
(-1)^{\epsilon_i+\epsilon_j}, \;\;\;\;\; \epsilon(S_0)=0.
\end{eqnarray}
The anti-Hamiltonian $S_0$ generates the vector fields
$\overleftarrow{U^a}$
\begin{eqnarray}
\label{Ua1} \overleftarrow{U^a}=(\;\cdot,S_0)^a =
-\overleftarrow{\nabla_i}\omega^{im}g_{mn}\theta^{na}(-1)^{\epsilon_m}-
\frac 12
\frac{\overleftarrow{\partial}}{\partial\theta_{ia}}\theta^m_c
(g_{mn}\nabla_i)\theta^{nc}(-1)^{\epsilon_n(\epsilon_i+1)+\epsilon_m}.
\end{eqnarray}

The conditions (\ref{mal}) yield the following equations for
$S_{0}$:
\begin{eqnarray}
S_0 U^a \equiv (S_0,S_0)^a=0, \qquad S_0V^a=0,\qquad S_0\Delta^a
=0. \label{S1}
\end{eqnarray}
In fact, these are equations to be fulfilled for $g_{ij}$. Of
course, solutions of these equations always exist. For example,
the covariant constant tensor field $g_{ij}$, $g_{ij}\nabla
_{k}=0$, belongs to them. According to this, the tensor field
$g_{ij}$ could be interpreted as a metric on $M$, thus making it
to a Riemannian manifold which, due to (\ref{R=0}), occurs to be
flat. However, more general, we do not restrict ourselves to this
special case, and just assume that equations (\ref{S1}) are
fulfilled.

In this way, we were able to generalize the construction of the
modified triplectic algebra, which in Ref.~\cite{gln} only was
given for an ordinary base manifold, to the case of a
supersymmetric base manifold. With that explicit realization of
the modified triplectic algebra on the supermanifold $\mathcal{M}$
we obtained all the ingredients for the quantization of general
gauge theories within the modified triplectic scheme.

In order to formulate this quantization procedure one repeats all
the essential steps, having performed for the first time in
Ref.~\cite{gln}, but now for the supermanifold $\mathcal{M}$. This
leads to the vacuum functional
\begin{eqnarray}
\label{Z}
%\nonumber
Z=\int dz \; d\lambda  \;{\cal D}_0\;{\rm exp}
{\{(i/\hbar)[W+X+\alpha S_0]\}},
\end{eqnarray}
where the quantum action $W=W(z)$ and the gauge fixing functional $%
X=X(z,\lambda )$ satisfy the following quantum master equations:
\begin{eqnarray}
\label{MEW} \hbox{\Large$\frac{1}{2}$}(W,W)^a +W{\cal V}^a=i\hbar
W\Delta^a ,
\end{eqnarray}
\begin{eqnarray}
\label{MEX} \hbox{\Large$\frac{1}{2}$}(X,X)^a +X{\cal U}^a=i\hbar
X\Delta^a.
\end{eqnarray}
Here, $\mathcal{D}_{0}$ is the integration measure,
\begin{eqnarray}
\label{mesu}
{\cal D}_0=({\rm sdet}[\omega^{ij}])^{-3/2},
\end{eqnarray}
$\alpha $ is an arbitrary constant, and the function $S_{0}$ has
been introduced in (\ref{S0}). In (\ref{MEW}) and (\ref{MEX}), we
have introduced generalized operators $\mathcal{V}^{a}$,
$\;\mathcal{U}^{a}$, according to
\begin{eqnarray}
\label{VU} {\cal V}^a=\hbox{\Large$\frac{1}{2}$}(\alpha U^a+\beta
V^a+\gamma U^a),\quad {\cal U}^a=\hbox{\Large$\frac{1}{2}$}(\alpha
U^a -\beta V^a-\gamma U^a).
\end{eqnarray}
Evidently, for arbitrary constants $\alpha $, $\beta $, $\gamma $ the
operators $\mathcal{V}^{a}$, $\mathcal{U}^{a}$ obey the properties
\begin{eqnarray}
{\cal V}^{\{a}{\cal V}^{b\}}=0,\quad
{\cal U}^{\{a}{\cal U}^{b\}}=0,\quad
{\cal V}^{\{a}{\cal U}^{b\}}+{\cal U}^{\{a}{\cal V}^{b\}}=0.
\end{eqnarray}
Therefore, the operators $\Delta ^{a}$, $\mathcal{V}^{a}$, $\mathcal{U}%
^{a} $ also realize the modified triplectic algebra.

The integrand of the vacuum functional (\ref{Z}) is invariant under the
BRST-antiBRST transformations defined by the generators
\begin{eqnarray}
\label{BRSTg}
 \delta^a=(\;\cdot, X-W)^a +{\cal V}^a-{\cal U}^a.
\end{eqnarray}

In the usual manner, one can prove that the vacuum functional (\ref{Z}), for
every fixed set of parameters $\alpha $, $\beta $, $\gamma $, does not
depend on the gauge-fixing function $X$.

\section{Conclusion}

In this paper, we have proposed a formulation of the modified triplectic
quantization in general coordinates.

We have found an explicit realization of the modified triplectic algebra of
generating operators $\Delta ^{a}$, $V^{a}$, $U^{a}$ on a triplectic
superspace $\mathcal{M}$, where the base supermanifold $M$ is a flat Fedosov
superspace endowed with a symmetric structure. The proposed scheme is
characterized by three free parameters, $\alpha$, $\beta$, $\gamma$,
whose specific choice, together with the Darboux coordinates, reproduces all
the known schemes of covariant quantization based on the BRST--antiBRST
invariance (for details, see \cite{gln}). Every specific choice of these
parameters $\alpha $, $\beta$, $\gamma$ gives a gauge-independent vacuum
functional and, therefore, a gauge independent S-matrix (see \cite{t}).\\
\\
{\sc Acknowledgements:}~The authors are grateful to D.V. Vassilevich for
stimulating discussions. The work was supported by Deutsche
Forschungsgemeinschaft (DFG), grant GE 696/7-1. The work of P.M.L. was also
supported by the projects INTAS 99-0590, DFG 436 RUS 113/669, by the
Russian Foundation for Basic Research (RFBR), 02-02-04002, 03-02-16193
and by the Pressident grant 1252.2003.2 for supporting leading scientific
schools.

%\noindent{\Large{\bf APPENDIX}}

\begin{appendix}

\section{Poisson and symplectic structures on supermanifolds}
\renewcommand{\theequation}{\thesection.\arabic{equation}}
\setcounter{equation}{0}
\hspace*{\parindent}
%\vspace{0.5cm}

Let us consider a second-rank tensor field
$\omega^{ij}=\omega^{ij}(x)$ of type $(2,0)$,
$\,\epsilon(\omega^{ij})=\epsilon(\omega)+\epsilon_i+\epsilon_j$,
which is defined on a supermanifold $M$ with local coordinates
$x^i,\, \epsilon(x^i)=\epsilon_i$. Then, let us introduce the
bilinear operation
\begin{eqnarray}
\label{A1}
(A,B)=\frac{\partial_r A}{\partial x^i}\;
(-1)^{\epsilon(\omega)\epsilon_i}\omega^{ij}\;
\frac{\partial B}{\partial x^j}=
\frac{\partial_r A}{\partial x^i}\;
 {^i\omega^j}\;
\frac{\partial B}{\partial x^j}\,,
\end{eqnarray}
which is invariant, $({\bar A},{\bar B})=(A,B)$, under arbitrary
coordinate transformations, $(x)\rightarrow ({\bar x})$.
Evidently, this operation obeys the following properties:
\begin{itemize}
\item[(a)] Grassmann parity
\begin{eqnarray}
\label{A2}
\epsilon(A,B)=\epsilon(A)+\epsilon(B)+\epsilon(\omega),
\end{eqnarray}
\item[(b)] linearity
\begin{eqnarray}
\label{A3}
&&(A+C,B)=(A,B)+(C,B),\quad (\epsilon(A)=\epsilon(C),\\
\nonumber &&(A,B+D)=(A,B)+(A,D),\quad \epsilon(B)=\epsilon(D)),
\end{eqnarray}
\item[(c)] Leibniz rule
\begin{eqnarray}
\label{A4}
\nonumber
(AC,B)&=&A(C,B)+(A,B)C(-1)^{\epsilon(C)(\epsilon(B)+\epsilon(\omega))},\\
(A,BD)&=&(A,B)D+B(A,D)(-1)^{\epsilon(B)(\epsilon(A)+\epsilon(\omega))}.
\end{eqnarray}
\end{itemize}
If, in addition, the tensor field $\omega^{ij}$ obeys the property
of generalized antisymmetry,
\begin{eqnarray}
\label{A5}
\omega^{ij}=-(-1)^{\epsilon_i\epsilon_j+\epsilon(\omega)}\omega^{ji}
\;\;\leftrightarrow \;\;
{^i\omega^j}=
-(-1)^{(\epsilon_i+\epsilon(\omega))
(\epsilon_j+\epsilon(\omega))}{^j\omega^i},
\end{eqnarray}
then, the binary operation (\ref{A1}) has the property of
\begin{itemize}
\item[(d)] generalized antisymmetry
\begin{eqnarray}
\label{A6} (A,B)=-
(-1)^{(\epsilon(A)+\epsilon(\omega))(\epsilon(B)+\epsilon(\omega))}(B,A).
\end{eqnarray}
\end{itemize}
Finally, let us restrict $\omega^{ij}$ to obey
\begin{eqnarray}
\label{A7} \omega^{ik}\partial_k\omega^{jn}
(-1)^{(\epsilon_i+\epsilon(\omega))(\epsilon_n+\epsilon(\omega))}
+ {\rm cycle} (i,j,n)\equiv 0,
\end{eqnarray}
then the operation (\ref{A1}) obeys
\begin{itemize}
\item[(e)] generalized Jacobi identity
\begin{eqnarray}
\label{A8}
(A,(B,C))(-1)^{(\epsilon(A)+\epsilon(\omega))(\epsilon(C)+\epsilon(\omega))}
+ {\rm cycle} (A,B,C)\equiv 0.
\end{eqnarray}
\end{itemize}
From (\ref{A2}), (\ref{A3}), (\ref{A4}), (\ref{A6}) and (\ref{A8}),
it follows that the operation (\ref{A1}) coincides with the
Poisson bracket when $\omega$ is even, $\epsilon(\omega)=0$, and
with the antibracket when $\omega$ is odd, $\epsilon(\omega)=1$.

Now, suppose that the tensor field $\omega^{ij}$ is non-degenerate.
Then, defining the right and left inverse matrices, labelled by $(R)$
and $(L)$, respectively, according to
\begin{eqnarray}
\label{A9}
{^i\omega^k}\;{_k\overset{(R)}{\omega}_j}={^i\delta_j},\quad
{_i\overset{(L)}{\omega}_k}\;{^k\omega^j}={_i\delta^j},
\end{eqnarray}
we find that these inverse matrices coincide:
\begin{eqnarray}
\label{A10}
{_i\overset{(R)}{\omega}_j}={_i\overset{(L)}{\omega}_j}={_i\omega_j},\quad
\epsilon({_i\omega_j})= \epsilon(\omega)+\epsilon_i+\epsilon_j\,.
\end{eqnarray}
Due to (\ref{A5}), we find that the inverse matrix ${_i\omega_j}$ has
the following property of generalized symmetry:
\begin{eqnarray}
\label{A11} {_i\omega_j}=
-(-1)^{\epsilon_i\epsilon_j+(\epsilon(\omega)
+1)(\epsilon_i+\epsilon_j)} {_j\omega_i}\quad\leftrightarrow\quad
\omega_{ij}=-(-1)^{\epsilon_i\epsilon_j}\omega_{ji}.
\end{eqnarray}
We also observe the remarkable fact that the symmetry properties
of the inverse matrix $\omega_{ij}$ do not depend on the Grassmann
parity of the tensor field $\omega$. Using the tensor field
$\omega_{ij}$ and DeWitt's index shifting rules (\ref{shift}), one
can rewrite the relations (\ref{A9}) in the forms
\begin{eqnarray}
\label{A}
\omega^{ik}\;\omega_{kj}(-1)^{\epsilon_k}=\delta^i_j,\quad
(-1)^{\epsilon_i}\omega_{ik}\;\omega^{kj}=\delta^j_i,
\end{eqnarray}
when $\epsilon(\omega)=0$, and
\begin{eqnarray}
\label{A0}
(-1)^{\epsilon_i}\omega^{ik}\;\omega_{kj}=\delta^i_j,\quad
\omega_{ik}\;\omega^{kj}(-1)^{\epsilon_k}=\delta^j_i,
\end{eqnarray}
in the case $\epsilon(\omega)=1$.
In terms of $\omega_{ij}$, the generalized Jacobi identity
(\ref{A7}) can be rewritten in the form
\begin{eqnarray}
\label{A12}
\partial_i\omega_{jk}(-1)^{\epsilon_i(\epsilon(\omega)+1+\epsilon_k)}
+ {\rm cycle} (i,j,k)\equiv 0 \;\;\leftrightarrow\;\;
\omega_{ij,k}(-1)^{\epsilon_i\epsilon_k}+ {\rm cycle} (i,j,k)
\equiv 0.
\end{eqnarray}

Let us now introduce a differential 2-form $\omega$ on the
supermanifold $M$, having the same form in both the even and
odd cases:
\begin{eqnarray}
\label{A13} \omega = \omega_{ij}dx^j\wedge dx^i, \quad dx^i\wedge
dx^j = -(-1)^{\epsilon_i\epsilon_j}dx^j\wedge dx^i.
\end{eqnarray}
It is invariant under a change of the local coordinates,
${\bar\omega}=\omega$. The external derivative of this 2-form is
given by
\begin{eqnarray}
\label{A14} d\omega = \omega_{ij,k}dx^k\wedge dx^j\wedge dx^i,
\quad d^2\omega=0.
\end{eqnarray}
It is also invariant under a change of the local coordinates,
$d{\bar \omega} = d\omega$. The requirement of closure
$(d\omega=0)$ leads exactly to the Jacobi identities for
$\omega_{ij}$ (\ref{A12}). Therefore, as in the case of the usual
differential geometry, there exists a one-to-one correspondence
between even (odd) non-degenerate Poisson supermanifolds and even
(odd) symplectic supermanifolds.

\section{Algebra of operators $\Delta^a$ and $V^a$}
\renewcommand{\theequation}{\thesection.\arabic{equation}}
\setcounter{equation}{0} \hspace*{\parindent}
%\vspace{0.5cm}

Let us investigate the algebra of the operators
$\overleftarrow{\Delta^a}$ (\ref{Delt}) and $\overleftarrow{V^a}$
(\ref{Va}). Omitting the details of the tedious calculations for
operators acting on scalars on ${\cal M}$, we obtain the following
results:
\begin{eqnarray}
\label{B1}
 \Delta^{\{a}\Delta^{b\}}&=&-
 \frac{\overleftarrow{\partial}}{\partial\theta_{ia}}
 \frac{\overleftarrow{\partial}}{\partial\theta_{jb}}
 \frac{\overleftarrow{\partial}}{\partial\theta_{mc}}\theta_{nc}
R^n_{\;\;mij}(-1)^{\epsilon_j(\epsilon_i+1)},\\
\label{B2}
 V^{\{a}V^{b\}}&=&-\frac 14
\frac{\overleftarrow{\partial}}{\partial\theta_{nc}}
\theta_{lc}R^l_{\;\;nim}\theta^{ia}\theta^{mb}
(-1)^{\epsilon_m(\epsilon_i+1)},\\
\label{B3}
%\nonumber
2(\Delta^aV^b+V^b\Delta^a)&=&
 \varepsilon^{ab}\frac{\overleftarrow{\partial}}{\partial\theta_{mc}}
\theta_{nc}R^n_{\;\;mij}\omega^{ij}\\
\nonumber && -
\frac{\overleftarrow{\partial}}{\partial\theta_{ia}}
\frac{\overleftarrow{\partial}}{\partial\theta_{mc}}\theta_{nc}
R^n_{\;\;mij}\theta^{jb}(-1)^{\epsilon_i+\epsilon_j}\\
 \nonumber &&
 - \varepsilon^{ab}\overleftarrow{\nabla}_j
\left(\omega^{ji}_{\;\;\;,i}
+\hbox{\Large$\frac{1}{2}$}\omega^{ji}\rho_{,i}\right)(-1)^{\epsilon_i} \\
\nonumber
&&+\frac{\overleftarrow{\partial}}{\partial\theta_{ma}}
\Big[-\Gamma^n_{\;\;mj.n}
(-1)^{\epsilon_j\epsilon_n+\epsilon_n(\epsilon_m+1)}+
\Gamma^n_{\;\;js}\Gamma^s_{\;\;mn}
(-1)^{\epsilon_j(\epsilon_m+\epsilon_n)+\epsilon_n(\epsilon_m+1)}\\
\nonumber &&
+\hbox{\Large$\frac{1}{2}$}\rho_{,mj}(-1)^{\epsilon_m}-
 \hbox{\Large$\frac{1}{2}$}\rho_{,i}\Gamma^i_{\;\;mj}(-1)^{\epsilon_m}\Big]
\theta^{jb}(-1)^{\epsilon_j}.
\end{eqnarray}

Let us introduce a function $\rho$, using the  relations
\begin{eqnarray}
    \label{B4} \omega^{ji}_{\;\;\;,i}  (-1)^{\epsilon_i}
    +\hbox{\Large$\frac{1}{2}$}\omega^{ji}\rho_{,i}  (-1)^{\epsilon_i}
    =0,
\end{eqnarray}
which, in view of (\ref{A}), is equivalent to
\begin{eqnarray}
\label{B5}
\rho_{,m}=2\omega^{ji}_{\;\;\;,i}\;\omega_{jm}
(-1)^{\epsilon_i+\epsilon_j}.
\end{eqnarray}
To solve these equations, it is necessary to use the consequences
of the Jacobi identities in the form
\begin{eqnarray}
\label{B6}
\omega_{ij}\omega^{ji}_{\;\;\;,m}+2\omega^{ji}_{\;\;\;,i}\;\omega_{jm}
(-1)^{\epsilon_i+\epsilon_j}=0.
\end{eqnarray}
Therefore, the function $\rho$ must satisfy the relations
\begin{eqnarray}
\label{B7}
\rho_{,m}=-\omega_{ij}\omega^{ji}_{\;\;\;,m}
\end{eqnarray}
and can be chosen as
\begin{eqnarray}
\label{B8}
 \rho=-{\rm log}\;{\rm sdet}\;[\omega^{ij}].
\end{eqnarray}
Indeed, for the variation of $\rho$ (\ref{B8}) we have
\begin{eqnarray}
\label{B9}
\nonumber
\delta\rho&=&-{\rm log}\;{\rm sdet}\;[\omega^{ij}+\delta\omega^{ij}]
+{\rm log}\;{\rm sdet}\;[\omega^{ij}]\\
\nonumber &=&-{\rm log}\;{\rm sdet}\;[\delta^i_j+
(-1)^{\epsilon_i}\omega_{il}\delta\omega^{lj}]\\
&=&-{\rm str}[(-1)^{\epsilon_i}\omega_{il}\delta\omega^{lj}]=
-\omega_{ij}\delta\omega^{ji}.
\end{eqnarray}
With allowance for (\ref{covom}) and (\ref{B5}), one can derive
the following useful relation:
\begin{eqnarray}
\label{B10} \hbox{\Large$\frac{1}{2}$}\rho_{,m}=
\Gamma^i_{\;\;mi}(-1)^{\epsilon_i(\epsilon_m+1)}=
\Gamma^i_{\;\;im} (-1)^{\epsilon_i}.
\end{eqnarray}

From (\ref{R}), (\ref{B4}) and (\ref{B10}) one gets the final
expression for the anticommutator (\ref{B3}):
\begin{eqnarray}
%\label{DaVb1}
%\nonumber
 \Delta^aV^b+V^b\Delta^a&=&\frac 12\Bigg(
\varepsilon^{ab}\frac{\overleftarrow{\partial}}{\partial\theta_{mc}}
\theta_{nc}R^n_{\;\;mij}\omega^{ij} -
\frac{\overleftarrow{\partial}}{\partial\theta_{ia}}
\frac{\overleftarrow{\partial}}{\partial\theta_{mc}}\theta_{nc}
R^n_{\;\;mij}\theta^{jb}(-1)^{\epsilon_i+\epsilon_j}\\
\nonumber
&&-\frac{\overleftarrow{\partial}}{\partial\theta_{ma}}
R^n_{\;\;mnj}\theta^{jb}(-1)^{\epsilon_n(\epsilon_m+1)+
\epsilon_j}\Bigg).
\end{eqnarray}
Taking this into account, we can see that when the base
supermanifod $M$ is a flat Fedosov superspace, $R^i_{\;\;jkm}=0$,
then on the triplectic supermanifold ${\cal M}$ we have an
explicit realization of the triplectic algebra (\ref{dd}) --
(\ref{vd}). Simultaneously, the operations (\ref{2op}) satisfy the
Jacobi identity (\ref{jiab}), and therefore (\ref{2op}) are
identified with extended antibrackets.

\end{appendix}


\begin{thebibliography}{99}
\bibitem{gln}  B. Geyer, P. Lavrov and A. Nersessian, Phys. Lett.
\textbf{B512} (2001) 211; Int. J. Mod. Phys. \textbf{A17} (2002) 1183.

\bibitem{bv}  I.A. Batalin and G.A. Vilkovisky, Phys. Lett. \textbf{%
B102} (1981) 27; Phys. Rev. \textbf{D28} (1983) 2567.

\bibitem{brst}  B.L. Voronov, and I.V. Tyutin, Theor. Math. Phys. %
\textbf{50} (1982) 218; \textit{ibid.} \textbf{52} (1982) 628. B.L.
Voronov, P.M. Lavrov and I.V. Tyutin, Yad. Fiz. \textbf{36} (1982) 498.
\newline
P.M. Lavrov, I.V. Tyutin, Russ. Phys. J. \textbf{25} (1982) 639;
Yad. Fiz. \textbf{41} (1985) 1685; Sov. J. Nucl. Phys.
\textbf{50} (1989) 912.\newline
P.M. Lavrov, P.Yu. Moshin and A.A. Reshetnyak, Mod. Phys. Lett.
\textbf{A10} (1995) 2687; JETP Lett. \textbf{62} (1995) 780.%
\newline
I.A. Batalin, K. Bering and P.H. Damgaard, Nucl. Phys. \textbf{B515}
(1998) 455.\newline
M.A. Grigoriev and P.H. Damgaard, Phys. Lett. \textbf{B474} (2000) 323.
\newline
P.M. Lavrov and P.Yu. Moshin, Theor. Math. Phys. \textbf{126} (2001) 101.

%BRST-antiBRST

\bibitem{BLT}  I.A. Batalin, P.M. Lavrov and I.V. Tyutin, J. Math. Phys.
\textbf{31} (1990) 1487; \textit{ibid.} \textbf{32} (1990) 532 (1990);
\textit{ibid.} \textbf{32} (1990) 2513.\\
P.M. Lavrov, Mod. Phys. Lett. \textbf{A6} (1991) 2051; Theor. Math.
Phys. \textbf{89} (1991) 1187.

\bibitem{3pl}  I.A. Batalin and R. Marnelius, Phys. Lett. \textbf{%
B350} (1995) 44; Nucl. Phys. \textbf{B465} (1996) 521;\newline
I.A. Batalin, R. Marnelius and A.M. Semikhatov, Nucl. Phys. \textbf{%
B446} (1995) 249.

\bibitem{gs1} M.A. Grigoriev and A.M. Semikhatov,
Phys. Lett. \textbf{B 417} (1998) 259,
\bibitem{gs2} M.A. Grigoriev and A.M. Semikhatov, Theor. Math. Phys.
\textbf{124} (2000) 1157.

\bibitem{anti}  P.M. Lavrov, Phys. Lett. \textbf{B366} (1996) 160;
Theor. Math. Phys. \textbf{107} (1996) 602.\newline
P.M. Lavrov and P.Yu. Moshin, Phys. Lett. \textbf{B508} 127 (2001);
Theor. Math. Phys. \textbf{129} (2001) 1645.

\bibitem{mod3pl}  B. Geyer, D.M. Gitman and P.M. Lavrov, Mod. Phys.
Lett. \textbf{A14} (1999) 661; Theor. Math. Phys. \textbf{123}
(2000) 813.

\bibitem{w}  E. Witten, Mod. Phys. Lett. \textbf{A5} (1990) 487.

\bibitem{geom}  O.M. Khudaverdian, J. Math. Phys. \textbf{32} (1991) 1934.
\newline
O.M. Khudaverdian and A.P. Nersessian, Mod. Phys. Lett. \textbf{A8}
 (1993) 2377; J. Math. Phys. \textbf{37} (1996) 3713. \newline
A. Schwarz, Commun. Math. Phys. \textbf{155} (1993) 249; \textit{%
ibid.} \textbf{158} (1993) 373. \newline
I.A. Batalin and I.V. Tyutin, Int. J. Mod. Phys. \textbf{A8} (1993) 2333;
Mod. Phys. Lett. \textbf{A8} (1993) 3673; \textit{ibid.}
\textbf{A9} (1994) 1707. \newline
H. Hata and B. Zwiebach, Ann. Phys.(N.Y.) \textbf{322} (1994) 131;
Phys. Lett. \textbf{B320} (1994) 29.
\newline
J. Alfaro and P.H. Damgaard, Phys. Lett. \textbf{B334} (1994) 369.%
\newline
O.M. Khudaverdian, Commun. Math. Phys. \textbf{198} (1998) 591;
\textit{Laplacians in Odd Symplectic Geometry}, [arXiv:math.DG/0212354].

\bibitem{DeWitt}  B. DeWitt, \textit{Supermanifolds}, Cambridge University
Press, Cambridge, 1984.

\bibitem{F}  B.V. Fedosov, J. Diff. Geom. \textbf{40} (1994) 213.%
\newline
B.V. Fedosov, \textit{Deformation quantization and index theory}, Akademie
Verlag (Berlin 1996).

\bibitem{fm}  I.~Gelfand, V.~Retakh and M.~Shubin, Advan. Math.
\textbf{136} (1998) 104; [dg-ga/9707024].

\bibitem{t}  I.V. Tyutin, Phys. Atom. Nucl. \textbf{65} (2002) 194.

\end{thebibliography}
\end{document}